\documentclass[graybox]{svmult}

\usepackage{mathptmx}       
\usepackage{helvet}         
\usepackage{courier}        
\usepackage{type1cm}        
\usepackage{makeidx}         
\usepackage{graphicx}        
\usepackage{multicol}        
\usepackage[bottom]{footmisc}

\makeindex             

\begin{document}

\title*{The SONG project and the prototype node at Tenerife}
\author{K. Uytterhoeven, P.L. Pall\'e, F. Grundahl, S. Frandsen, H. Kjeldsen, J. Christensen-Dalsgaard, M. Fredslund Andersen, E. Weiss, U. G. J$\o$rgensen, P.K. Rasmussen, A.N. S$\o$rensen, K. Harps$\o$e, \& J. Skottfelt}
\authorrunning{Uytterhoeven et al.}
\institute{K. Uytterhoeven \& P.L. Pall\'e \at 1)  Instituto de Astrof\'{\i}sica de Canarias (IAC), Tenerife; 2) Dept. Astrof\'{\i}sica, Universidad de La Laguna (ULL), Tenerife \email{katrien@iac.es}
\and F. Grundahl, S. Frandsen, H. Kjeldsen, J. Christensen-Dalsgaard, M. Fredslund Andersen, \& E. Weiss \at Department of Physics and Astronomy, Aarhus University  
\and U.G. J$\o$rgensen, P.K. Rasmussen, A.N. S$\o$rensen, K. Harps$\o$e, \& J. Skottfelt \at Niels Bohr Institute and Centre for Star and Planet Formation, University of Copenhagen 
}
%
%
\maketitle

\abstract*{SONG (Stellar Observations Network Group) is a global network of 1-m class robotic telescopes that is under development. The SONG prototype will shortly be operational at Observatorio del Teide, Tenerife, and first light is expected by December 2011. The main scientific goals of the SONG project are asteroseismology of bright stars and follow-up and characterization of exo-planets by means of precise measurements of stellar surface motions and brightness variations. We present the Tenerife SONG node and its instruments.}

\abstract{We present the SONG network and its prototype node at Teide Observatory.}

\section{The SONG project}
\label{sec:1}
SONG  (Stellar Observations Network Group) is an initiative to construct a global network of 1-m robotic telescopes to ensure continuous monitoring (\cite{2006},  \cite{2009}). The main scientific goals of the SONG project are asteroseismology of bright stars and follow-up and characterization of exoplanets by means of precise measurements of stellar surface motions and brightness variations. 

Two instruments are foreseen on a SONG node: (1) a dual-colour lucky imaging camera; (2) a high-resolution (R$\sim$120,000) \'echelle spectrograph with wavelength coverage  4700-7000 \AA\, and with a radial velocity precison down to $<$2m/s. 

The SONG network aims eventually at 8 nodes. The prototype SONG node is currently under construction at Observatorio del Teide, Tenerife (SONG-OT). Funding is approved for a second node at Delingha Observatory, China, with first light in 2013. A US proposal is in preparation for the funding of a US node, and a Danish proposal is currently under evaluation for funding of additional nodes.

\section{The prototype SONG node}
\begin{figure}
\begin{tabular}{ccc}
\includegraphics[width=40mm,height=30mm]{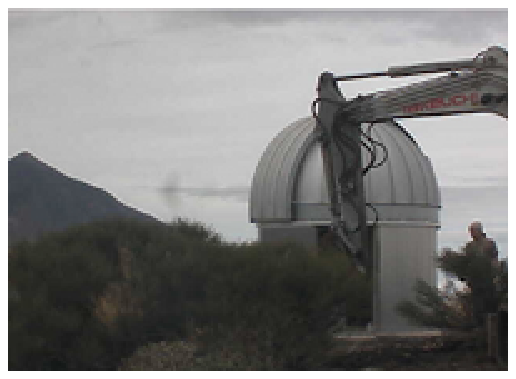}
&
\includegraphics[width=40mm,height=30mm]{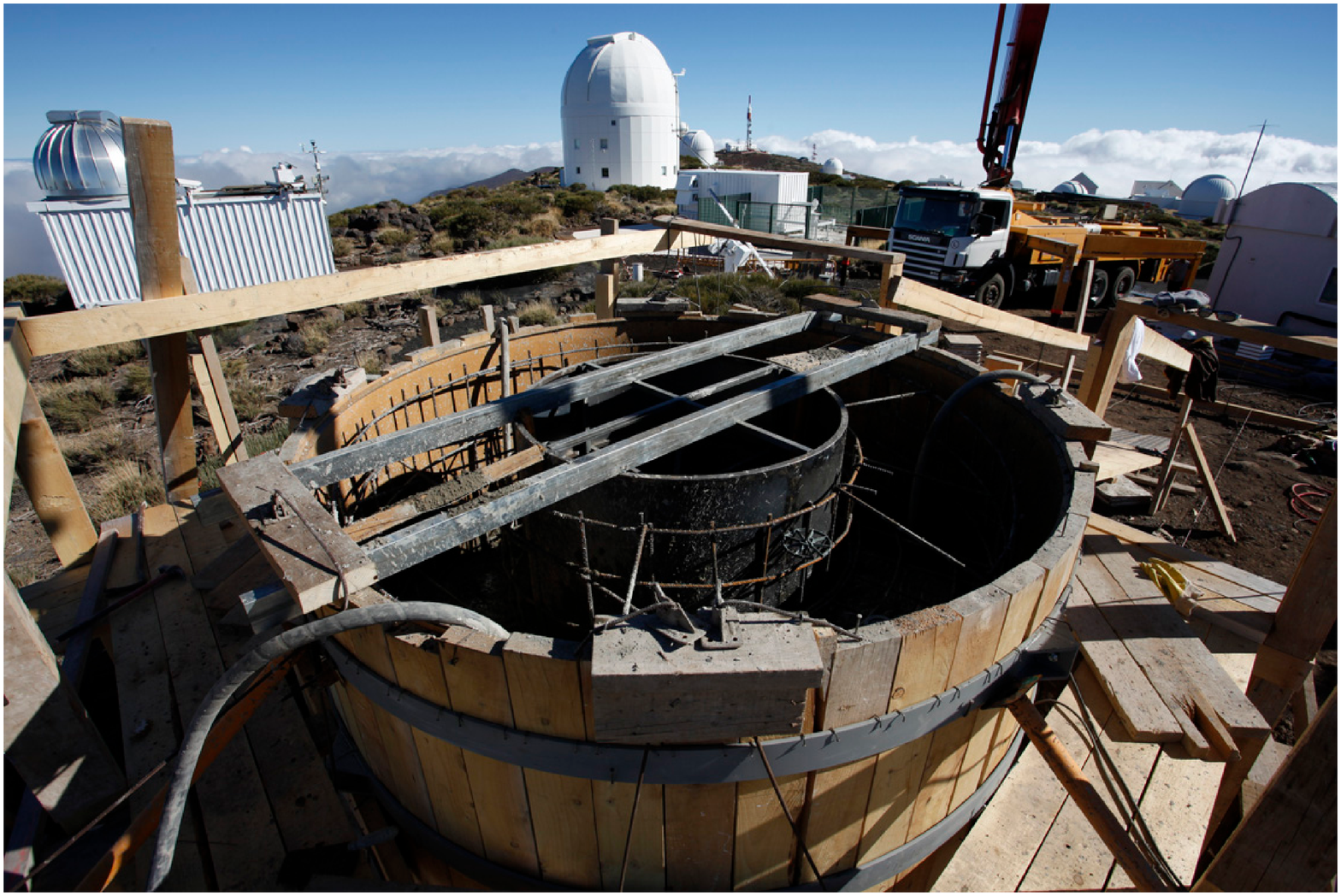}
&
\includegraphics[width=40mm,height=30mm]{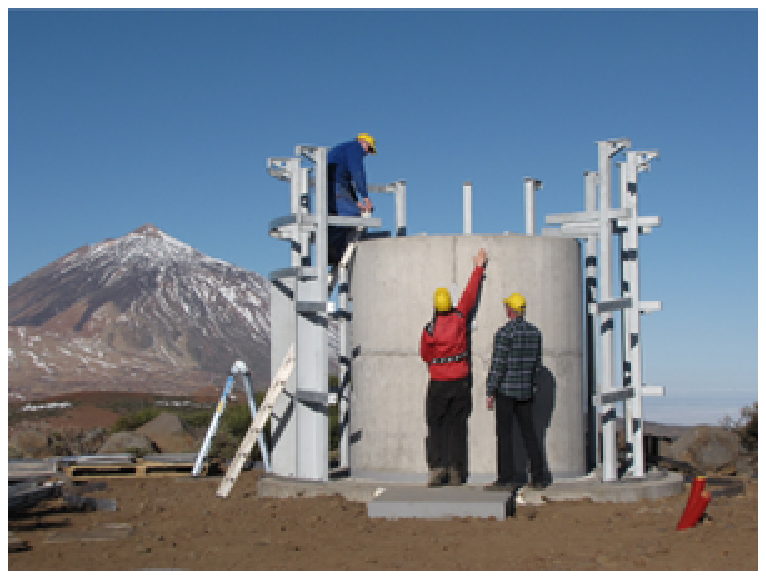}
\\
\includegraphics[width=40mm,height=30mm]{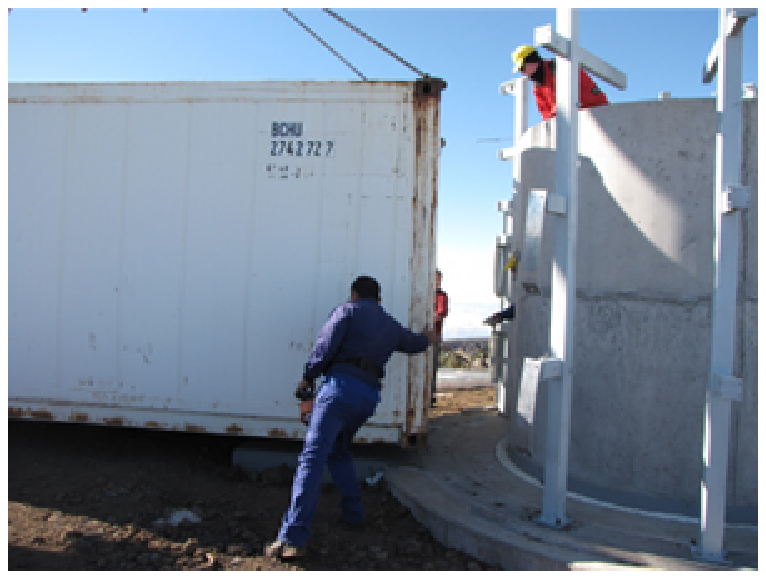}
&
\includegraphics[width=40mm,height=30mm]{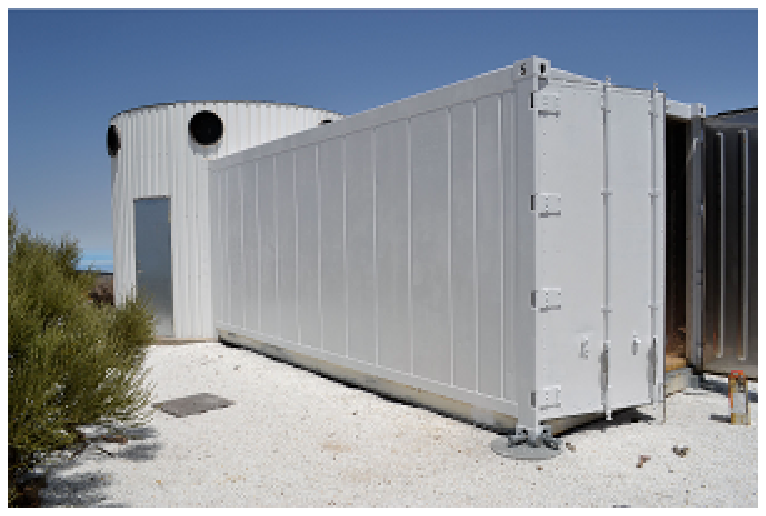}
&
\includegraphics[width=40mm,height=30mm]{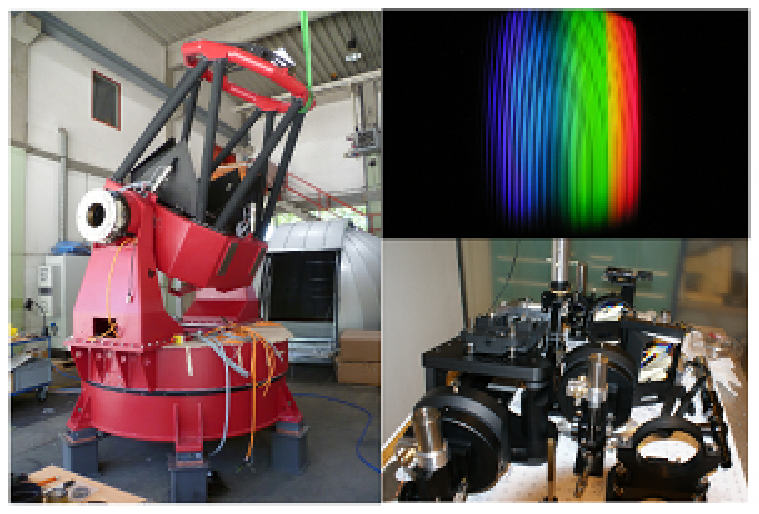}
\end{tabular}
\caption{The evolution of SONG-OT between October 2010 and September 2011 (timeline from left to right, from top to bottom): 
the removal of the STARE telescope, the construction of the dome support and container, the telescope, the spectrograph, and an example of the \'echelle spectrum.}
\label{fig:1}       
\end{figure}

First light of the prototype SONG node, SONG-OT, is foreseen by the end of 2011. The evolution and current status of SONG-OT is illustrated in Figure\,\ref{fig:1}. 

{\bf Building:} The removal of the STARE telescope, excavation and pouring of the concrete SONG pier took place in October/November 2010. The SONG container and dome support were installed in March 2011. \\
{\bf Telescope and dome:} Factory acceptance is scheduled for the end of October 2011. Installation is expected in November/December 2011. \\
{\bf Spectrograph and Lucky Imaging Cameras:} The instruments are developed and tested in Aarhus and Copenhagen, and are waiting for first light.

%

\begin{acknowledgement}
We thank the Villum Foundation, the Carlsberg Foundation and The Danish Council for Independent Research | Natural Sciences for support for the SONG prototype. KU acknowledges financial support by the Spanish National Plan of R\&D for 2010, project AYA2010-17803. 
\end{acknowledgement}

\end{document}